\documentclass[prl,reprint,superscriptaddress]{revtex4-1}

\usepackage{amsmath,graphicx,textcomp}

\usepackage{import}
\usepackage{siunitx}

\makeatletter
\renewcommand{\@seccntformat}[1]{}
\makeatother

\begin{document}

%\preprint{}

\title{Model-free measurement of the pair potential in colloidal fluids using optical microscopy}

% Please use the appropriate macro for the type of \textbf{information}


\author{Adam Edward \surname{Stones}}
\email{adam.stones@chem.ox.ac.uk}

\affiliation{Department of Chemistry, Physical \& Theoretical Chemistry Laboratory, University of Oxford, Oxford OX1 3QZ, United Kingdom}
\author{Roel P. A. \surname{Dullens}}
\affiliation{Department of Chemistry, Physical \& Theoretical Chemistry Laboratory, University of Oxford, Oxford OX1 3QZ, United Kingdom}
\author{Dirk G. A. L. \surname{Aarts}}
\email{dirk.aarts@chem.ox.ac.uk}
%\thanks{}
\affiliation{Department of Chemistry, Physical \& Theoretical Chemistry Laboratory, University of Oxford, Oxford OX1 3QZ, United Kingdom}


\date{\today}

\begin{abstract}
\noindent We report a straightforward, model-free approach for measuring pair potentials from particle-coordinate data, based on enforcing consistency between the pair distribution function measured separately by the distance-histogram and test-particle insertion routes. We demonstrate the method's accuracy and versatility in simulations of simple fluids, before applying it to an experimental system composed of superparamagnetic colloidal particles. The method will enable experimental investigations into many-body interactions and allow for effective coarse-graining of interactions from simulations.
\end{abstract}

%\keywords{}

%\pacs{89.20.Kk}

\maketitle %\maketitle must follow title, authors, abstract and \pacs

\noindent The versatility of colloidal particles as model systems for condensed matter physics can hardly be overstated, as illustrated through studies in crystallisation \cite{Tan2014,Feng2015}, glasses \cite{Schall2007,Mattsson2009}, gels \cite{Lu2008}, interfacial phenomena \cite{Aarts2004}, and rheology \cite{Cheng2011}. Moreover, the colloidal interactions can be engineered to give rise to novel self-assembled phases \cite{Sacanna2010,Chen2011} with exciting potential applications. Despite these tremendous successes for colloid science, \textit{quantitative} comparisons between experiments and theory or simulations, and thus more accurate and powerful predictions from those theoretical tools, hinge on precise knowledge of the interparticle interactions. These interactions are often only approximately known and are based on theoretical assumptions, for example that the particles interact through a DLVO or depletion potential \cite{Israelachvili2015}.

Existing methods to measure the pair potential come with certain drawbacks: methods based on inverting correlation functions rely on additional assumptions \cite{Carbajal1996,Rajagopalan1997,Behrens2001,Quesada2001} or large numbers of computer simulations \cite{Schommers1973,Levesque1985,Brunner2002,Royall2003}, and direct methods \cite{Crocker1994,Piech2002,Hertlein2008} are typically carried out away from the equilibrium context, measured only between two particles or a particle and a wall. Furthermore, these methods can be technically demanding, both in the lab and numerically, and are therefore not routinely employed. By contrast, in this report we provide a versatile, fast, model-free approach which allows an effective pair potential to be determined solely from particle coordinates obtained using optical microscopy, based on a novel inversion of the pair distribution function. The method is broadly applicable in fields dealing with the liquid state, ranging from soft matter to biophysics, as well as in a wide range of industries where colloidal systems are used.

The pair distribution function $g$ offers a real-space visualisation of pairwise structure. It provides a direct link to the phase behaviour and thermodynamic properties of the system if the interparticle interactions are known \cite{Hansen2013}. For a homogeneous fluid, $g$ is given by the ratio of the local number density about a reference particle, $\rho^{(0)}(\mathbf{r})$, and the bulk number density, $\rho$:
\begin{equation}
	g(\mathbf{r}) = \frac{\rho^{(0)}(\mathbf{r})}{\rho}\mathrm{,}
	\label{eq:PDFDensity}
\end{equation}
where $\mathbf{r}$ is the position vector relative to the reference particle. In experiments and simulations, $\rho^{(0)}(\mathbf{r})$ is typically measured using a distance-histogram method \cite{Allen2017}.

Here, we propose an alternative approach based on test-particle insertion \cite{Henderson1983,Stones2018,Widom1963}, with $g$ again given by a ratio of local and bulk ensemble averages:
\begin{equation}
	g(\mathbf{r}) = \frac{\langle\exp(-\Psi/k_{\mathrm{B}}T)\rangle^{(0)}_{\mathbf{r}}}{\langle\exp(-\Psi/k_{\mathrm{B}}T)\rangle}\mathrm{,}
	\label{eq:PDFInsertion}
\end{equation}
in which $\Psi$ is the additional potential energy due to the hypothetical insertion of a particle and $k_{\mathrm{B}}T$ is the thermal energy. The potential energy $\Psi$ is written as a sum of pairwise interactions, depending on the effective pair potential $u$ between the particles in the fluid. For a homogeneous system at a given density, $u$ giving rise to a particular $g$ is unique \cite{Henderson1974}. Crucially, matching $g$ from insertion with that from the distance-histogram method provides an elegant route to obtain the pair potential $u$.

We will first explain the method, before demonstrating its use and accuracy in simulation. We then apply the method to a colloidal model system composed of superparamagnetic particles, in which the pairwise interaction can be tuned using an external magnetic field \cite{Promislow1995}. The experimental demonstration is for a quasi-two dimensional one-component fluid composed of particles with isotropic interactions, but we stress that equation (\ref{eq:PDFInsertion}) holds for anisotropic interactions in any dimension and may be extended to multi-component fluids.

%%% Figure 1 %%%%%%%%%%%%%
\begin{figure*}[]
    \includegraphics[width=\textwidth]{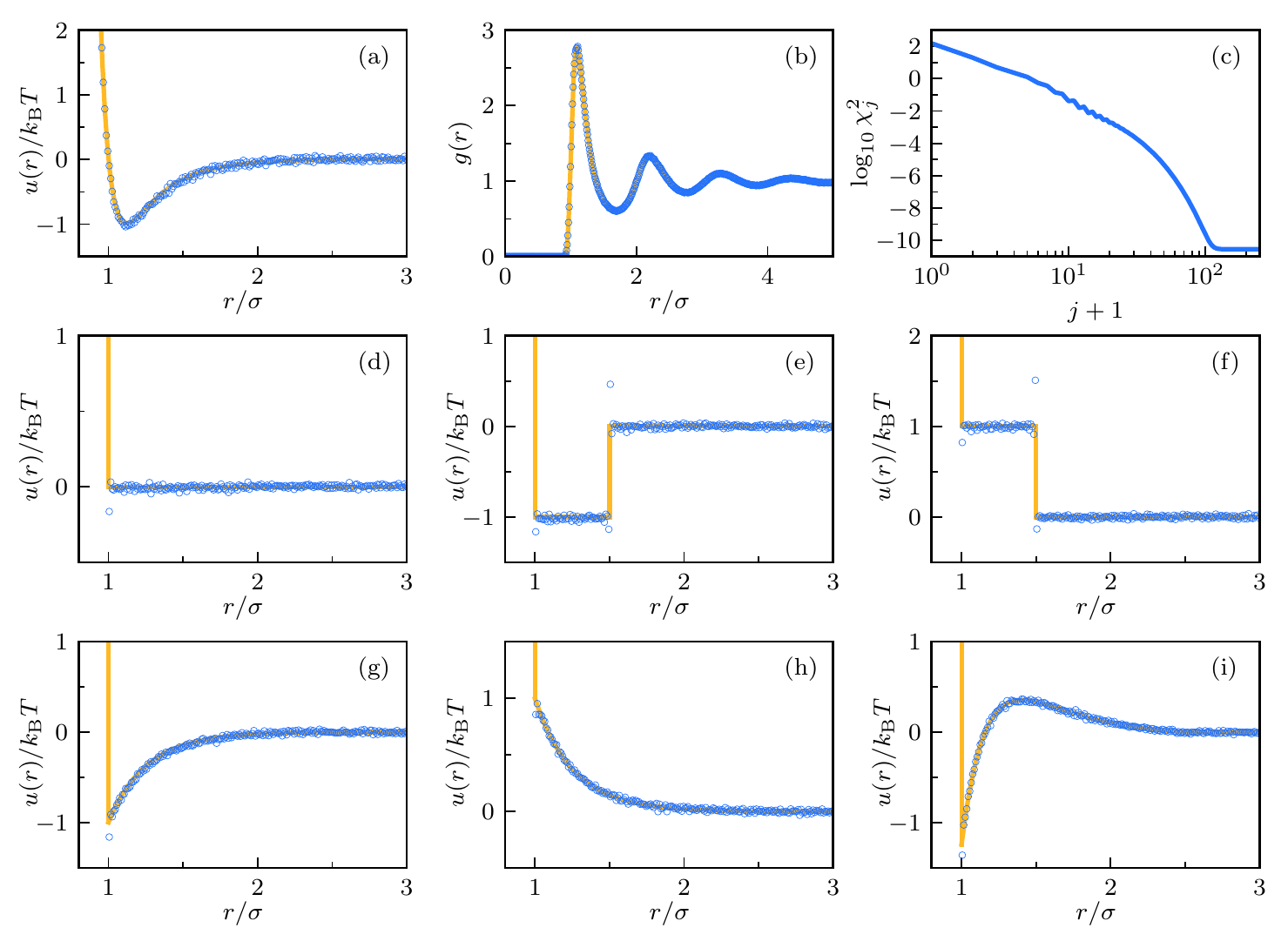}
    \caption{(a) Lennard-Jones $u(r)$ measured using the method presented here (blue circles), compared with the simulation input (yellow line). (b) Corresponding $g(r)$ obtained by the distance-histogram method (yellow line) and by test-particle insertion with the measured pair potential (blue circles). (c) Log-log plot showing convergence of the PC scheme. Remaining panels show pair potentials measured for (d) a hard-disk potential, (e) a hard-core attractive square-well potential, (f) a hard-core repulsive square-well potential, (g) an attractive Yukawa potential, (h) a repulsive Yukawa potential, (i) a hard-core two-Yukawa (HCTY) potential.}
    \label{fig:simulationAll}
\end{figure*}
%%%%%%%%%%%%%%%%%%%%%%%%%%

%%% Figure 2 %%%%%%%%%%%%%
\begin{figure*}[]
    \includegraphics[width=\textwidth]{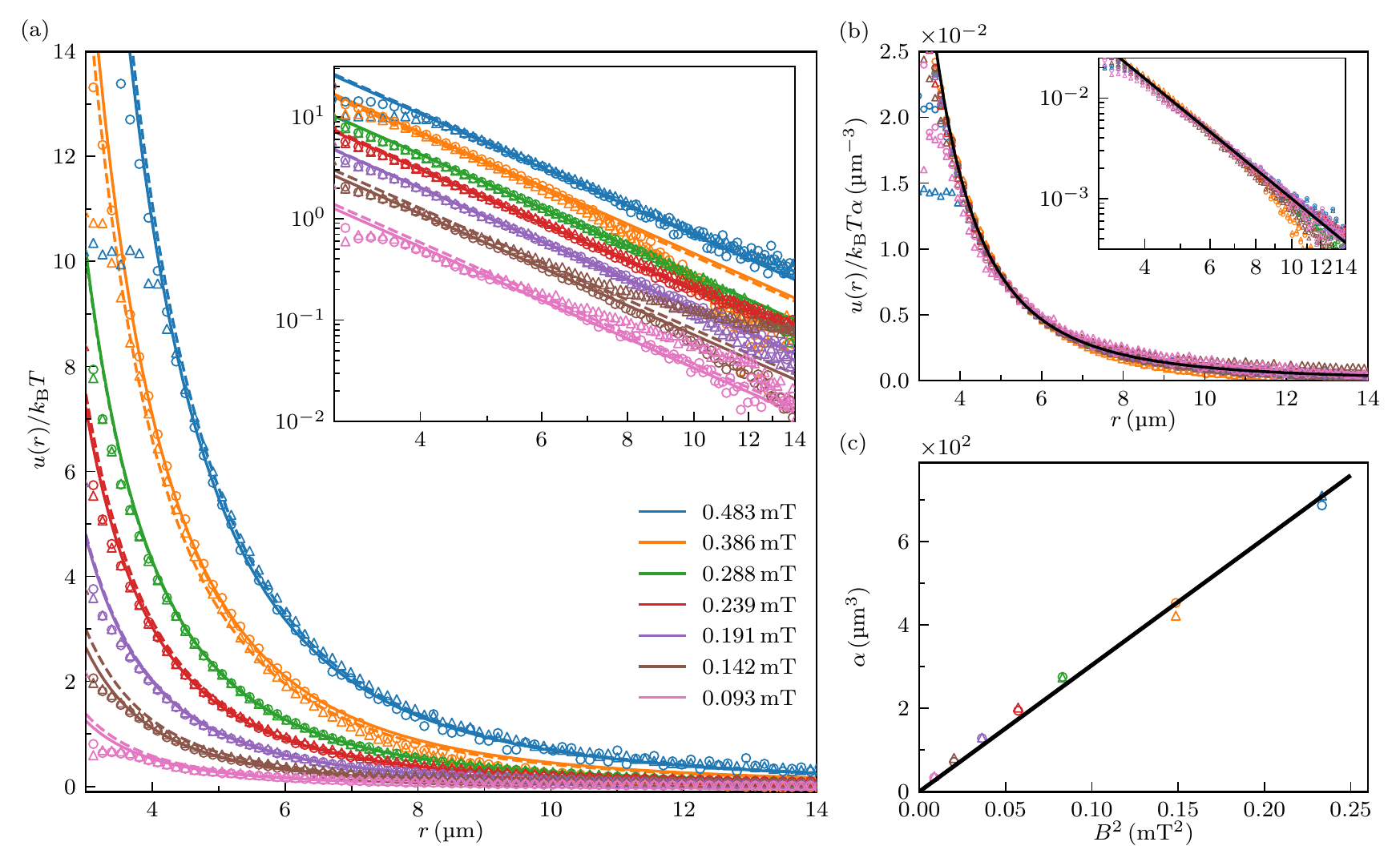}
    \caption{(a) Repulsive dipolar pair potentials measured for two samples at different number densities (Sample 1---triangles and dashed lines, Sample 2---circles and solid lines) in \emph{experiment} for different $B$; the lines are fits to (\ref{eq:DipolarPairPotential}). Inset: corresponding log-log plot where the pair potentials appear as parallel straight lines with slope $-3$. (b) When divided by the fitting parameter $\alpha$, the measured pair potentials collapse onto a single curve. (c) Plot of $\alpha$ against $B^2$---as anticipated, the data fall on a straight line.}
    \label{fig:experiment}
\end{figure*}
%%%%%%%%%%%%%%%%%%%%%%%%%%

To find $u$, a predictor-corrector (PC) scheme proposed by Schommers \cite{Schommers1973} is used. Briefly, a trial pair potential $u_{j}(r)$---here, we begin with $u_{0}(r)=0$ everywhere---is used in equation (\ref{eq:PDFInsertion}) to obtain a prediction of the pair distribution function $g_{j}(r)$. This is compared with the distance-histogram result $g_{\mathrm{h}}(r)$ using
\begin{equation}
    u_{j+1}(r) = u_{j}(r) - k_{\mathrm{B}}T \ln\left(\frac{g_{\mathrm{h}}(r)}{g_{j}(r)}\right)\mathrm{,}
    \label{eq:SchommersCorrector}
\end{equation}
to obtain a corrected pair potential $u_{j+1}(r)$. This scheme is performed iteratively until convergence, which is verified by monitoring the value of 
\begin{equation}
    \chi_{j}^2 = \sum_{k=0}^{k=N} (g_{j}(r_{k})-g_{\mathrm{h}}(r_{k}))^2\mathrm{,}
    \label{eq:ChiSquared}
\end{equation}
where the $r_k$ are the mid-points of the bins used in the distance-histogram method. In contrast with inverse Monte Carlo schemes \cite{Schommers1973,Levesque1985}, the calculation of $g_{j}(r)$ required for the corrector step (\ref{eq:SchommersCorrector}) is performed using test-particle insertion on the \emph{existing} particle coordinates and each iteration does not require an additional (expensive) simulation. The distances between each test-particle and those of the fluid need to be calculated only once, allowing for a very efficient implementation. A more detailed account of the implementation of this PC scheme is provided in the supplemental material \cite{supplementalMaterial}.

The scheme was first tested using particle coordinates generated by Monte Carlo simulations, where the measured pair potentials can be compared to the simulation input \cite{supplementalMaterial}. Figure \ref{fig:simulationAll}(a-c) demonstrates the use of the method with data from a two-dimensional Lennard-Jones simulation. The agreement between the measured and input pair potentials is excellent. As expected, the PC scheme enforces consistency between the two methods, converging after $\sim\!100$ iterations (see Figure \ref{fig:simulationAll}(c))---in the other cases, the convergence is even faster.

Figures \ref{fig:simulationAll}(d-i) show the application of the method to a wide range of pair potentials of interest in liquid-state theory and colloid science, including hard disks \cite{Thorneywork2017}, attractive and repulsive square-well potentials, attractive and repulsive hard-core Yukawa potentials, and a hard-core two-Yukawa potential \cite{Rudisill1989} with competing short-range attractions and long-range repulsions---in all cases the input pair potential is recovered. We particularly note the discontinuities in the input pair potentials, which are captured by our analysis. 

Next, we apply the method to an experimental colloidal model system, where the pair potential is generally unknown, illustrating the strength of our method. We used a quasi two-dimensional system of superparamagnetic colloidal particles with a diameter $\sigma\approx\SI{3}{\micro\metre}$, which acquire a magnetic dipole moment when placed in an external magnetic field \cite{supplementalMaterial}. Outside of the core region ($r>\sigma$), the particles are \emph{expected} to interact with a repulsive dipolar pair potential
\begin{equation}
    \frac{u(r)}{k_\mathrm{B}T} = \frac{\alpha}{r^3}\mathrm{,}
    \label{eq:DipolarPairPotential}
\end{equation}
with $\alpha$ proportional to the square of the magnetic flux density $B$ \cite{Promislow1995}. This system is particularly useful for this investigation since several different $u(r)$ can be measured using a single sample by altering $B$.

The measured pair potentials are shown in Figure \ref{fig:experiment}(a) for two different particle number densities---in all cases, a pair potential is readily extracted showing the expected inverse-cube decay, as illustrated on the log-log plot of the inset. Equation (\ref{eq:DipolarPairPotential}) can be used to extract a value of $\alpha$ for each $u(r)$, with larger $B$ inducing a larger dipole moment in each particle, corresponding to a stronger dipolar repulsion. The measured pair potentials collapse onto a single curve after dividing by the values of $\alpha$ found from the fit, as shown in Figure \ref{fig:experiment}(b). In Figure \ref{fig:experiment}(c), the extracted values of $\alpha$ are plotted against $B^2$, yielding the anticipated straight line. Finally, we note that for a given $B$, the same pair potential is obtained in both samples, showing that at these densities the system is pairwise additive. 

In summary, we have demonstrated a novel method for measuring pair potentials in colloidal systems. The method is readily extended to three-dimensional and multi-component systems, and will be a valuable tool in characterising colloidal particles used in fundamental and applied studies. The experimental and computational requirements are minimal: a simple transmission light microscope to obtain the particle snapshots and a desktop computer will suffice. Standard image analysis routines may be used to obtain the particle coordinates and the computational scheme can be straightforwardly encoded and run. We stress that no assumptions about the form of the pair potential have to be made. In addition to clear industrial applications, where knowledge of interparticle interactions is often essential in formulation of products, the method will allow for fundamental investigations into many-body interactions, such as those arising in the controversy around like-charge attractions \cite{Grier1998}, and enables coarse-grained pair potentials \cite{Lyubartsev1995} to be derived for use in multi-scale simulations of more complex fluids and biological systems.

\begin{acknowledgments}

A.E.S. gratefully acknowledges financial support from the University of Oxford Clarendon Fund. Carla Fern\'{a}ndez-Rico is thanked for useful discussions and practical assistance.

\end{acknowledgments}

\bibliography{pair_potentials}

\end{document}

%\preprint{}

\title{Supplemental material: Model-free measurement of the pair potential in colloidal fluids using optical microscopy}

% Please use the appropriate macro for the type of \textbf{information}


\author{Adam Edward \surname{Stones}}
\email{adam.stones@chem.ox.ac.uk}

\affiliation{Department of Chemistry, Physical \& Theoretical Chemistry Laboratory, University of Oxford, Oxford OX1 3QZ, United Kingdom}
\author{Roel P. A. \surname{Dullens}}
\affiliation{Department of Chemistry, Physical \& Theoretical Chemistry Laboratory, University of Oxford, Oxford OX1 3QZ, United Kingdom}
\author{Dirk G. A. L. \surname{Aarts}}
\email{dirk.aarts@chem.ox.ac.uk}
%\thanks{}
\affiliation{Department of Chemistry, Physical \& Theoretical Chemistry Laboratory, University of Oxford, Oxford OX1 3QZ, United Kingdom}


\date{\today}

%\keywords{}

%\pacs{89.20.Kk}

\maketitle %\maketitle must follow title, authors, abstract and \pacs

\section{Simulations}

\textbf{Protocol} We implemented a Grand Canonical Monte Carlo Scheme \cite{Frenkel2001}, in which the number of particles $N$ is allowed to vary according to a fixed chemical potential $\mu$.

\textbf{Parameters} Each simulation was carried out in a two-dimensional square box of length $50$ with periodic boundary conditions. The particle-size parameter $\sigma$ was set to $1$ in each simulation. The configuration space was explored using three trial moves: particle insertion, particle deletion and particle movement, in the ratio $3:1:1$. Initially, $10$ particles were placed in the box at random and the density was allowed to increase during $10^6$ equilibration steps. Subsequently, particle coordinates were taken every $10^4$ moves, giving a total of 1000 snapshots. Details of the input pair potentials and parameters are given in Table \ref{table:simulationDetails}. In all cases, the pair potential was truncated at $2.5\sigma$, and shifted such that $u(2.5\sigma)=0$ to maintain continuity.

{
\renewcommand{\arraystretch}{1.4}
 \begin{table*}[]
     \caption{The input pair potentials and parameters used in the simulation. Also shown is the mean density of each simulation.}
     \label{table:simulationDetails}
 \begin{ruledtabular}
 \begin{tabular}{p{2cm} l l c c}
     Potential                       & $u(r)$  & Parameters & $\mu$ & $\overline{\rho\sigma^2}$ \\ \hline \\
     Lennard-Jones                   & $ 4\left(\left(\frac{r}{\sigma}\right)^{12}-\left(\frac{r}{\sigma}\right)^6\right)$ & $\begin{array}{ll}
                                                                                                                                    \sigma   &= 1 \\
                                                                                                                                    \end{array}     $ & $1$ & 0.677  \\\\
     Hard disk                       & $
                                        \begin{array}{l@{\quad}l}
                                        \infty    &\ r\leq\sigma\\
                                        0         &\ r>\sigma\\
                                        \end{array} $                                                                     & $\begin{array}{ll}
                                                                                                                                    \sigma   &= 1 \\
                                                                                                                                    \end{array} $  & $2$ & 0.481  \\\\
     Attractive square-well          & $
                                        \begin{array}{l@{\quad}l}
                                        \infty    &\ r\leq\sigma\\
                                        \epsilon &\ \sigma<r\leq\lambda\sigma\\
                                        0         &\ r>\lambda\sigma\\
                                        \end{array} $                                                                     & $
                                                                                                                                    \begin{array}{ll}
                                                                                                                                    \sigma   &= 1 \\
                                                                                                                                    \lambda  &= 1.5 \\
                                                                                                                                    \epsilon &= -1 \\
                                                                                                                                    \end{array}  $  & $-1$  & 0.539 \\\\
     Repulsive square-well          & $
                                        \begin{array}{l@{\quad}l}
                                        \infty    &\ r\leq\sigma\\
                                        \epsilon &\ \sigma<r\leq\lambda\sigma\\
                                        0         &\ r>\lambda\sigma\\
                                        \end{array} $                                                                     & $
                                                                                                                                    \begin{array}{ll}
                                                                                                                                    \sigma   &= 1 \\
                                                                                                                                    \lambda  &= 1.5 \\
                                                                                                                                    \epsilon &= 1 \\
                                                                                                                                    \end{array}  $  & $1$ & 0.264  \\\\
     Attractive Yukawa          & $
                                        \begin{array}{l@{\quad}l}
                                        \infty    &\ r\leq\sigma\\
                                        \left(\frac{\epsilon \sigma}{r}\right)\exp\left(-\kappa\left(\frac{r}{\sigma}-1\right)\right) &\ r>\sigma\\
                                        \end{array} $                                                                     & $
                                                                                                                                    \begin{array}{ll}
                                                                                                                                    \sigma   &= 1 \\
                                                                                                                                    \kappa  &= 1.5 \\
                                                                                                                                    \epsilon &= -1 \\
                                                                                                                                    \end{array}  $  & $1$ & 0.580 \\\\
     Repulsive Yukawa          & $
                                        \begin{array}{l@{\quad}l}
                                        \infty    &\ r\leq\sigma\\
                                        \left(\frac{\epsilon \sigma}{r}\right)\exp\left(-\kappa\left(\frac{r}{\sigma}-1\right)\right) &\ r>\sigma\\
                                        \end{array} $                                                                     & $
                                                                                                                                    \begin{array}{ll}
                                                                                                                                    \sigma   &= 1 \\
                                                                                                                                    \kappa  &= 1.5 \\
                                                                                                                                    \epsilon &= 1 \\
                                                                                                                                    \end{array}  $  & $1$ & 0.306 \\\\
         Hard-core two-Yukawa (HCTY) & $
                                        \begin{array}{l@{\quad}l}
                                        \infty    &\ r\leq\sigma\\
                                        \left(\frac{\epsilon \sigma}{r}\right)\left[-\exp\left(-z\left(\frac{r}{\sigma}-1\right)\right) + A\exp\left(-y\left(\frac{r}{\sigma}-1\right)\right)\right] &\ r>\sigma\\
                                        \end{array} $                                                                     & $
                                                                                                                                    \begin{array}{ll}
                                                                                                                                    \sigma   &= 1 \\
                                                                                                                                    \epsilon  &= 6 \\
                                                                                                                                    A &= 0.8 \\
                                                                                                                                    z &= 4 \\
                                                                                                                                    y &= 2.5\\
                                                                                                                                    \end{array}  $ & $1$ & 0.367 \\\\
 \end{tabular}
 \end{ruledtabular}
 \end{table*}
}

\section{Experiments}

\textbf{Protocol} Superparamagnetic spheres with a diameter $\sigma\approx\SI{3}{\micro\metre}$ (Dynabeads\textsuperscript{\tiny\textregistered} M-270 Carboxylic Acid, Invitrogen) in 20/80 \% \emph{v}/\emph{v} ethanol/water were allowed to sediment in a quartz glass cell (Hellma Analytics) to form a quasi-two dimensional colloidal monolayer. The particles were magnetised by applying a magnetic field perpendicular to the sample plane using a solenoid. Samples of two different number densities were used, and measurements were taken for a range of different magnetic fields. Images were taken every second using an Olympus CKX41 bright-field microscope fitted with a 40$\times$ objective and a Ximea XIQ CMOS camera. A summary of the experimental data used in the analysis is given in Table \ref{table:experimentDetails}.

{
\renewcommand{\arraystretch}{1.4}
 \begin{table}[]
     \caption{Summary of the experimental data used in the analysis.}
     \label{table:experimentDetails}
 \begin{ruledtabular}
 \begin{tabular}{c c c c}
     Sample & Field (mT)  & Frames & ~\,$\overline{\rho\sigma^2}$\footnote{We used a diameter of $\sigma=\SI{3.04}{\micro\metre}$, based on the extracted pair potentials. Note that the number density may vary within the sample due to fluctuations or changing the location of the field of view.}\\ \hline
     1 & 0.093 & 3600 & 0.139\\
     1 & 0.142 & 3600 & 0.138\\
     1 & 0.191 & 3600 & 0.143\\
     1 & 0.239 & 2865 & 0.144\\
     1 & 0.288 & 3600 & 0.144\\
     1 & 0.386 & 3600 & 0.143\\
     1 & 0.483 & 3600 & 0.110\\
     
     2 & 0.093 & 3600 & 0.175\\
     2 & 0.142 & 3600 & 0.179\\
     2 & 0.191 & 3600 & 0.176\\
     2 & 0.239 & 3600 & 0.177\\
     2 & 0.288 & 3600 & 0.180\\
     2 & 0.386 & 3600 & 0.192\\
     2 & 0.483 & 3600 & 0.177\\
 \end{tabular}
 \end{ruledtabular}
 \end{table}
}

\textbf{Interactions} Assuming the spheres are uniformly magnetised, they behave as point magnetic dipoles from the centre of the sphere \cite{Edwards2017}. The system is quasi-two dimensional and the induced magnetic dipole moments are perpendicular to the sample plane, and so assuming that the spheres are magnetically identical, the pair potential outside of the hard core is given by
\begin{equation}
    u(r) = \frac{\mu_0 m^2}{4\pi r^3}\mathrm{,}
    \label{eq:dipoleDipoleInteractionExperiment}
\end{equation}
where $m$ is the magnitude of the magnetic dipole moment on each particle and $\mu_0$ is the magnetic constant.

For sufficiently small fields, $m$ is proportional to the magnetic flux density $B$ and is given by $m=V_{\mathrm{p}} \xi B/\mu_0$ \cite{Ge2011}, where $\xi$ is the dimensionless volume susceptibility and $V_{\mathrm{p}} = \pi \sigma^3 /6$ is the particle volume. Substituting into (\ref{eq:dipoleDipoleInteractionExperiment}) and dividing by the thermal energy $k_{\mathrm{B}}T$ yields
\begin{equation}
    \frac{u(r)}{k_\mathrm{B}T} = \frac{\alpha}{r^3}\mathrm{,}
    \label{eq:DipolarPairPotential}
\end{equation}
with
\begin{equation}
    \alpha = \frac{\pi \sigma^6 \xi^2 B^2}{144 k_{\mathrm{B}}T\mu_0}\mathrm{.}
    \label{eq:alphaMagnetic}
\end{equation}

\textbf{Field calibration} A solenoid was attached to the microscope in order to apply a magnetic field perpendicular to the sample plane. We measured the resulting magnetic flux density $B$ at the centre of the solenoid (the sample location) using a Gaussmeter (GM07, Hirst Magnetic Instruments Ltd.).

\begin{figure}[]
    \includegraphics[width=\columnwidth]{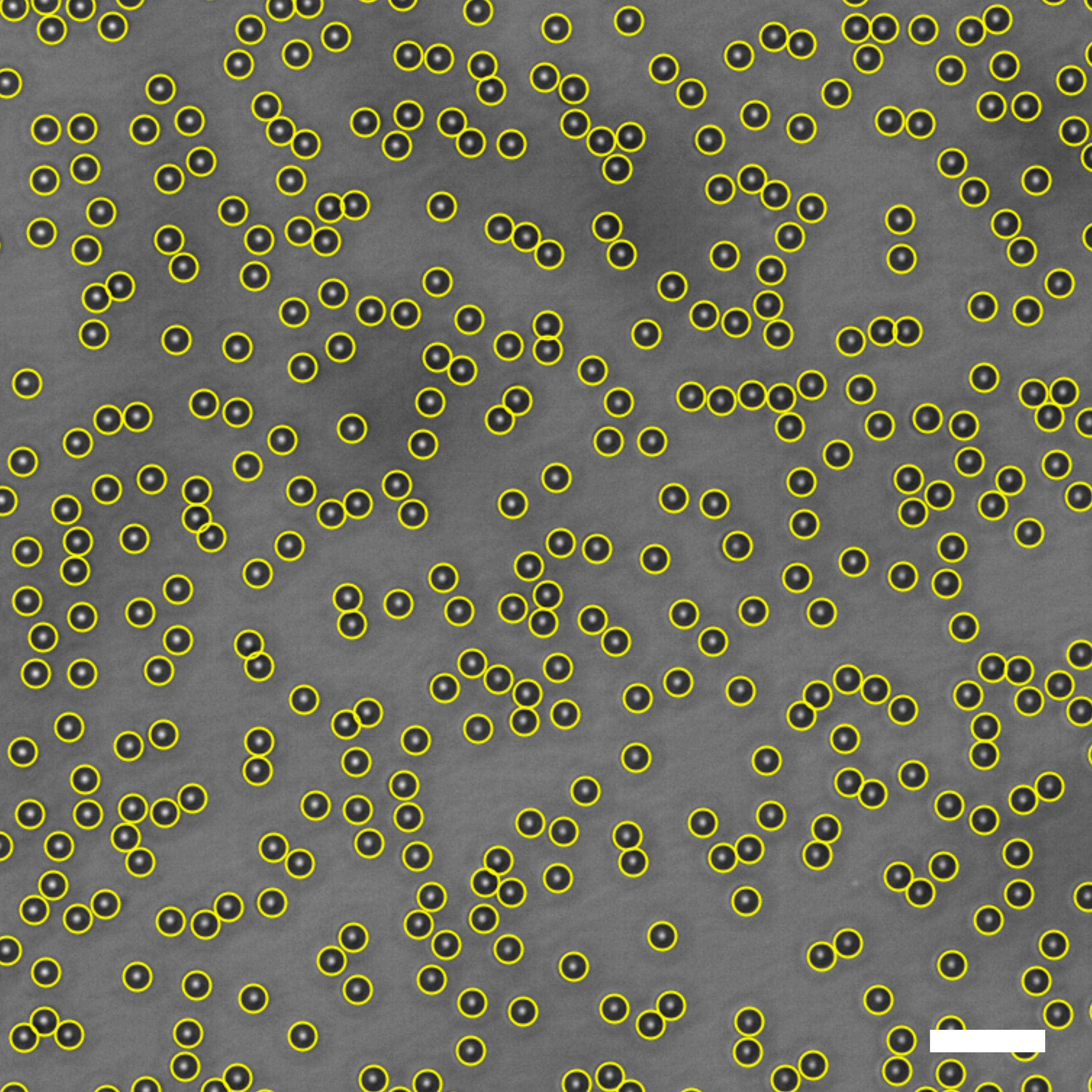}
    \caption{Part of a typical image with particles' positions superimposed. Scale bar corresponds to $\SI{15}{\micro\metre}$. Note that the image has been adjusted so that the pixel intensities are spread between $0$ and $1$. Taken from Sample 2 with $B=\SI{0.093}{\milli\tesla}$.} 
    \label{fig:image}
\end{figure}

\textbf{Image analysis} An example of an image with the particles' positions superimposed is shown in Figure \ref{fig:image}. The positions were measured using a customised algorithm which locates particles based on both the bright spot at the particle centre and the dark ring surrounding the particle. The images are first adjusted so that pixel intensities are spread between $0$ and $1$, and then each pixel is divided by the mean of all pixels within a certain radius to account for differences in the background throughout the images. 

To detect the dark rings, the adjusted images are negated, and their background removed by clipping at their median pixel intensity; the images are then binarised, eroded by one pixel and their edges are detected using a Sobel filter. The outer rings are then detected using a circular Hough transform, from which the accumulators are stored. The bright spots in the accumulators correspond with the best candidates for the centres of the dark rings.

To detect the particle centres, the background of the adjusted images is removed by clipping at their median pixel intensity; the bright spots in the resulting images correspond to the centres of the particles. These images are multiplied pixel-wise by the corresponding accumulators from the Hough transform. The local maxima of these products correspond to the particles; these are then detected and their coordinates are refined using standard routines \cite{Crocker1996,TrackPy}. Finally, spurious peaks corresponding to interstitial sites between particles are removed using distance criteria.

%%% Figure S1 %%%%%%%%%%%%%
%%%%%%%%%%%%%%%%%%%%%%%%%%

\section{Predictor-corrector scheme}

The distance-histogram method was used to measure $g_\mathrm{h}(r)$ at a range of distances ${r_k}$ corresponding to the centres of the histogram intervals. The insertion result $g_j(r)$ is then calculated for the trial pair potential $u_j(r)$. Although the result can be applied directly by attempting test-particle insertion at fixed distances from the particles in the system \cite{Stones2018}, it is more efficient to attempt insertion at points on a grid over the field of view or simulation box, and approximate $\langle\exp(-\Psi/k_{\mathrm{B}}T)\rangle^{(0)}_{r_k}$ for each $r_k$ by averaging over points falling in the same intervals used when calculating $g_\mathrm{h}(r)$. To calculate $\Psi$ for each point, we find the distances to neighbouring particles within a cut-off radius, assuming that beyond this distance the pair potential is zero. We then use the trial pair potential $u_j(r)$ to calculate $\Psi$ as a sum of pairwise interactions---a lookup table is used to accomplish this more efficiently.  Note that the distances between the insertion points and their neighbouring particles are calculated only once, contributing to the performance of the iterative procedure.

The insertion result $g_j(r)$ is subsequently used in (3) to obtain the next correction to the pair potential, $u_{j+1}(r)$. Numerically, care must be taken when using this equation, since $g_\mathrm{h}(r)$ and $g_j(r)$ for small values of $r_k$ are often zero, and their logarithms cannot be taken. We therefore recast the corrector (3) as
\begin{equation}
    e_{j+1}(r) = e_{j}(r)\frac{g_{\mathrm{h}}(r)}{g_{j}(r)}\mathrm{,}
    \label{eq:SchommersCorrectorExponential}
\end{equation}
where $e_{j}(r)=\exp(-u_{j}(r)/k_{\mathrm{B}}T)$. After calculating $g_{j}(r)$, values of zero are replaced by $10^{-20}$ to avoid dividing by zero. The resulting $e_{j+1}(r)$ will be zero where $g_\mathrm{h}(r)$ is zero, and so these values are also replaced by $10^{-20}$ before the pair potential is calculated as $u_{j+1}(r)=-k_{\mathrm{B}}T\ln e_{j+1}(r)$.

\section{Analysis}

\textbf{Simulation} All 1000 snapshots were analysed, with 10000 insertion points used in each. The cut-off for the pair potential was $5\sigma$, with a comparison interval of $10^{-2}\sigma$ and lookup interval of $5\times 10^{-3}\sigma$. In each case, 250 iterations of the PC scheme were performed, and convergence was achieved.

\textbf{Experiment} All of the frames indicated in Table \ref{table:experimentDetails} were used for the distance-histogram calculation, and every tenth frame was used for the insertion analysis, with 10000 insertion points in each. The cut-off for the pair potential was $\SI{34.6}{\micro\metre}$, with a comparison interval of $\SI{0.138}{\micro\metre}$ and lookup interval of $\SI{0.014}{\micro\metre}$. Because some particles at the edges of the images were missed by the detection algorithm, we used a border cut-off of $\SI{13.8}{\micro\metre}$. In each case, 500 iterations of the PC scheme were performed and the algorithm converged. Each pair potential was fit with a dipolar repulsion by recasting (\ref{eq:DipolarPairPotential}) as
\begin{equation}
    \ln\left(\frac{u(r)}{k_\mathrm{B}T}\right) = \ln{\alpha} - 3 \ln{r}\mathrm{,}
\end{equation}
and fitting the data to a straight line (corresponding to the log-log plot in Figure 2) to extract $\alpha$. In each case, the data between $\SI{3.94}{\micro\metre}$ and $\SI{7.13}{\micro\metre}$ was used, since the log-log plot shows the obtained pair potential corresponds well to a dipolar repulsion in this region.

The resulting values of $\alpha$ for each $B$ were then fitted using (\ref{eq:alphaMagnetic}), and a value of $\xi\sim0.9$ was extracted. Note that this value depends sensitively on the particle diameter ($\xi\propto\sigma^{-3}$); we used $\sigma=\SI{3.04}{\micro\metre}$ based on the extracted pair potentials. The temperature was estimated as $T=\SI{298}{\kelvin}$.
{
\makeatletter
\renewcommand{\@seccntformat}[1]{}
\makeatother

\section{Data and code availability}

The data and code used in this letter are available from the corresponding authors on request. All figures have associated raw data.

}
\bibliography{pair_potentials}